\documentclass[twocolumn,showpacs,preprintnumbers,amsmath,amssymb]{revtex4}

\usepackage{graphicx}
\usepackage{dcolumn}
\usepackage{bm}

\begin{document}

\title{A Zeeman Slower based on magnetic dipoles}

\author{Yuri B. Ovchinnikov}
\affiliation{National Physical Laboratory, Hampton Road, Teddington, Middlesex TW11 0LW, United Kingdom}

\date{\today}
\begin{abstract}
A transverse Zeeman slower composed of an array of compact
discrete neodymium magnets is considered. A simple and precise
model of such a slower based on magnetic dipoles is developed. The
theory of a general Zeeman slower is modified to include spatial
nonuniformity of the slowing laser beam intensity due to its
convergence and absorption by slowed atoms. The slower needs no
high currents or water cooling and the spatial distribution of its
magnetic field can be adjusted. In addition the slower provides a
possibility to cool the slowed atoms transversally along the whole
length of the slower. Such a slower would be ideal for
transportable optical atomic clocks and their future applications
in space physics.
\end{abstract}



\maketitle

\section{Introduction}
The sources of slow atoms based on laser cooling of their
translation degree of freedom are widely used in many modern
atomic physics experiments. At the present time a Zeeman slower
\cite{Phillips} is the most efficient slower of that type.
Recently a maximal flux of cold Rb atoms as high as $3.2 \times 10^{12}\,$at/s, produced
by a Zeeman slower with an additional section for the transverse laser cooling of atoms,
has been demonstrated \cite{Slowe}.
For some alkali atoms like Rb and Cs a Zeeman slower can
be substituted with more compact sources based on
magneto-optical traps \cite{Riis, Dieckmann, Ovchinnikov}, the
typical flux of which is between $10^9$ and $10^{10}\,$at/s. The most
intense source of that type based on two-dimensional magneto-optical
trap is able to produce the flux of $6\times10^{10}\,$at/s \cite{Schoser}.
On the other hand, for many other atoms and experimental conditions a Zeeman slower is the only choice.
A standard Zeeman slower uses a special
current-carrying coil to create a proper spatial distribution of the
magnetic field along its axis, the purpose of which is to compensate a
Doppler shift of the decelerated atoms with a corresponding Zeeman shift.

An advantage of a Zeeman slower based on permanent magnets is that
it does not need high currents and corresponding cooling of its
coils. The existent approach to such a permanent Zeeman slower
consists in substitution of standard Zeeman coils with analogous
ring-like magnets made of plastic-bonded permanent magnets \cite{Adler}, which
can be machined to any form prior their magnetization. The
disadvantage of these magnets is that they are not as strong as
dense magnets and their strength is subject to fluctuations due
to the inconsistancy of the density of the magnetic material.

In this paper it is proposed to use an array of small neodymium
magnets to create a required magnetic field distribution of a
Zeeman slower. A simple model of the slower, based on point-like magnetic
dipoles is introduced. It is shown that this model works well even for 
magnets of a size comparable to their distance from the axis of the slower.
The design of such a slower includes a possibility
to vary the distances of the individual magnets from the axis of
the slower, which allows fine tuning of the magnetic field.

In the first part of the paper the existent standard theory of a
Zeeman slower is extended to apply it to a cooling laser beam with
a nonuniform distribution of its intensity along the axis
of the slower. It is shown how the absorption of the cooling light
can be included in calculation of the most optimal spatial distribution
of the magnetic field of the slower. It is proposed in the case of a nonuniform
laser field to use instead of a standard design parameter $\eta$ a slightly different
parameter $\epsilon$, which relates the designed local deceleration of atoms
to its maximum possible value at the same location at given laser intensity.
It is shown that the most optimal cooling of the decelerated atoms takes place
at $\epsilon=0.75$.

In the first section a numerical approach to the calculation of
the optimal spatial distribution of the magnetic field of a
Zeeman slower is described for a general case of a cooling laser
beam with non-uniform intensity distribution. In the second
section Zeeman slowers based on permanent magnetic dipoles are
discussed. In the outlook section the construction of a transverse
Zeeman slower based on neodymium magnets and its applications are
discussed. Finally, in the conclusion the main results are
summarized.

\section{General theory of Zeeman slower}

In a Zeeman slower atoms are decelerated in a counterpropagating
resonant laser beam due to momentum transfer from spontaneously
scattered photons. An existent analytical theory of a Zeeman
slower \cite{Napolitano, Metcalf} is based on uniform deceleration
of atoms in a field of a laser beam with uniform intensity
distribution. In practice intensity of the slowing
laser beam is never uniform. There are several
reasons for that. First, to increase an overlap of the laser beam
with the expanding atomic beam and to provide some additional transverse
cooling of the decelerated atoms, convergent laser beam is usually used.
Second, absorption of laser light by slowed atoms essentially changes
the distribution of intensity of light along the axis of a Zeeman
slower.

Below a procedure of numerical calculation of the
magnetic field distribution of a Zeeman slower is described for a general case
of a slowing laser beam with non-uniform spatial distribution of
its intensity. To keep it simple, dependence of the
light intensity along only one of the coordinates z, which is
the axis of a Zeeman slower, is considered.

The spontaneous light pressure force is given by
\begin{equation}
F(v,z)=\frac{\hbar k \Gamma}{2} \frac{s_0(z)}{1+s_0(z)+4[\delta_0+kv-\mu' B(z)/\hbar]^2/\Gamma^2},
\end{equation}
where $k$ is the wave vector of light, $s_0(z)$ is the local
on-resonance saturation parameter of the atomic transition,
$\Gamma$ is the linewidth of the transition, $\delta_0$ is the
laser frequency detuning, $v$ is the velocity of the atom, $\mu'$ is
the magnetic moment for the atomic transition and $B(z)$ is the
local magnitude of the magnetic field. The velocity dependence
of the force is determined by the effective frequency detuning
$\Delta_{eff}=\delta_0+kv-\mu' B(z)/\hbar$,
which includes the Doppler shift $kv$ of the atomic frequency. The
maximum value of the local deceleration, provided by the force, is
achieved at exact resonance, when $\Delta_{eff}=0$, and is given by
\begin{equation}
a_{max}(z)=\frac{\hbar k \Gamma}{2 m} \frac{s_0(z)}{1+s_0(z)},
\end{equation}
where m is the mass of the atom. Although the maximum
deceleration can be used to estimate the shortest possible
length of a Zeeman slower, this can not be realised in practice. At
exact resonance an equilibrium between the inertial force of the
decelerated atoms and the light force is unstable and any
slight increase of the atomic velocity due to
imperfection of the magnetic field distribution or spontaneous
heating of atoms will lead to decrease of the decelerating force
and subsequent loss of atoms from the deceleration process. In
practice the deceleration of atoms is realized at a fraction of
the maximum deceleration
\begin{equation}
a(z)=\epsilon a_{max}(z),
\end{equation}
where $\epsilon<1$. Note that the coefficient $\epsilon$ in our
case corresponds to the ratio between the reduced local acceleration
and the maximum possible acceleration at the same location, which
is a function of the local saturation parameter $s_0(z)$. In a standard
treatment \cite{Napolitano, Metcalf} a similar coefficient
$\eta$ appears, which relates the actual deceleration to the maximum possible
deceleration at infinite intensity of laser light. When the deceleration is less than $a_{max}$ 
the decelerated atoms stay on the low-velocity wing
of the Lorentzian velocity profile of the light force (1) and andergo
stable deceleration. The corresponding equilibrium velocity $v(z)$
of the decelerated atoms, which is less than the
resonant velocity $v_{res}(z)=(\mu' B(z)/\hbar-\delta_0)/k$, is
determined by
\begin{equation}
kv(z)=kv_{res}(z)-\frac{\Gamma}{2} \sqrt{(1+s_0(z))\frac{1-\epsilon}{\epsilon}}.
\end{equation}

The most optimal offset of the equilibrium velocity $v(z)$ from
the resonant velocity $v_{res}(z)$ is achieved at a point where
the derivative of the force (1) reaches its maximum, because the
damping of the relative motion of atoms around this point is
maximal. It is easy to show that within our definition of the
$\epsilon$ coefficient (3) this most optimal cooling condition is
achieved exactly at $\epsilon=0.75$. The damping coefficient of
the force decreases with the increas of the intensity of the laser
field. Therefore, choosing of the most optimal intensity of the
cooling beam for a Zeeman slower is a compromise between having a
large deceleleration force and low velocity spread of the
decelerated atoms. Usually the saturation parameter of a Zeeman
slower is chosen to be about 1.

The procedure of calculation of the spatial distribution of a Zeeman slower
looks as follows. First, the actual velocity
of the slowing atoms is calculated numerically according to the formula
\begin{equation}
\frac{dv(z)}{dz}=\epsilon \frac{\hbar k \Gamma}{2 m} \frac{s_0(z)}{1+s_0(z)},
\end{equation}

After that the resonant velocity can be determined from the eq.\,(4) and the
corresponding distribution of the magnetic field calculated as
\begin{equation}
B(z)=\hbar (\delta_0+kv_{res}(z))/\mu'.
\end{equation}
The distribution of the saturation parameter $s_0(z)$ on the axis
of the slower has to be taken according to the convergence of the
cooling laser beam and its absorption by the slowed atoms. If the influence of the
convergence of the laser beam to the spatial
distribution of the saturation parameter is easy to include, but the calculation
of the absorption of light by the decelerated atoms is not so
straightforward. The problem of the propagation of atoms and light
through each other while they are mutually interacting is
difficult to solve. On the other hand, it can be easily solved by
inverting the direction of motion of the atoms and calculating their
acceleration along the laser beam, starting from the end edge of a
Zeeman slower.
For atoms copropagating with the laser beam, the change of the atomic velocity,
which is responsible for the local density of atoms, and the change
of the intensity of light at each point inside the slower can be simultaneously
computed from their values at the preceding spatial step. The
corresponding distribution of the saturation parameter along the Zeeman
slower can be found from the equation
\begin{equation}
\frac{d s_0(z^\ast)}{d z^\ast}=-s_0(z^\ast) n(z^\ast) \frac{ \sigma_0 \epsilon} {1+s_0(z^\ast)}+s_0(z^\ast)\frac{2}{l_0-z^\ast},
\end{equation}
where $z^\ast=z_f-z$ is a coordinate along the direction of
propagation of the laser beam, which starts at the end point
$z=z_f$ of the Zeeman slower, $n(z^\ast)$ is the local density of
atoms, $\sigma_0$ is the resonant light scattering cross
section and $l_0$ is the distance from the end edge of the slower to
the waist of the converged laser beam. The first term in the right side of the equation
is responsible for the absorption of light and the second one for the convergence of the laser beam.
It is supposed that the focal plane of the laser beam is located outside the slower, such as
$l_0$ is larger than its length.
In this one-dimensional problem the density
means the number of atoms per unit length along the $z$-axis of the slower.
The local density of the slowed atoms at $z^\ast$ is given by
\begin{equation}
n(z^\ast)=\frac{A}{v(z^\ast)} \int_{v(z^\ast )}^{v_0} \Phi(v_z,u) dv_z,
\end{equation}
where $\Phi(v_z,u)=0.5(v_z^3/u^4)\exp(-0.5 v_z^2/u^2)$ is the flux
density in the initial thermal atomic beam, $u=\sqrt{k_B T/m}$,
$v_0$ is the capture velocity of the slower and $A$ is a
coefficient to normalize the density to the total flux of
atoms. In this formula the density of atoms at the location $z^\ast$
is determined by the integrated flux of atoms between the
velocities $v(z^\ast)$ and $v_0$, which is divided by the local
velocity of the slowed atoms $v(z^\ast)$.
The analytical
solution of the eq. (8) is given by
\begin{equation}
n(z^\ast)=A \frac{(2u^2+v(z^\ast)^2)e^{-\frac{v(z^\ast)^2}{2 u^2}}-(2u^2+v_0^2)e^{-\frac{v_0^2}{2u^2}}}{2 u^2 v(z^\ast)}.
\end{equation}
This function predicts
rapid increase of the density of the slowed atoms towards the end
of the slower, where the velocity $v(z^\ast)$ becomes small,
which was confirmed experimentally in
\cite{Firmino, Napolitano} by direct measurement of the
fluorescence of atoms inside a Zeeman slower.

As an example we will consider here a Zeeman slower for Sr atoms
similar to one in \cite{Lemonde}, where the $^1S_0 \rightarrow
^1P_1$ transition of Sr with wavelength $\lambda=461\,$nm and
natural linewidth $\Gamma=2 \pi \times 32\,$MHz is used for
cooling of the translation motion of atoms. The slower is designed
to slow atoms from the initial velocity $v_0=420\,$m/s down to the
final velocity $25\,$m/s over a distance $L_s=25\,$cm with
the efficiency parameter $\epsilon=0.6$.
The value of the $\epsilon$ parameter was chosen to be below its optimal value $\epsilon=0.75$ to
include possible imperfection of the magnetic field distribution of the slower.
The thermal velocity of the
initial atomic beam is taken to be $u=292\,$m/s, which corresponds
to the temperature of $T=630\,^{\circ}$C. For the given capture
velocity of the slower, the flux of the slowed
atoms is about 28\% of the total initial flux of thermal atoms.
The cooling laser beam is taken to be convergent, such that its diameter
at the output of the slower ($z=z_f=25\,$cm) is $d_0=1\,$cm and at
its input ($z=0$) $d_1=0.3\,$cm. This corresponds to the distance
between the output end of the slower and the waist of the laser beam of
$l_0=L_sD_0/(d_0-d_1)=35.7\,$cm. The saturation
parameter at the end edge of the slower is set to be $s_0(z_f)=1$.
The total absorption of the cooling
light in the slower was set to 50\% of the total light power of the laser
beam. Taking into account that each decelerated atom absorbs in
average about 22000 photons it is easy to derive that absorption of 22.5\,mW of laser power
gives the total flux of cold atoms
about $2.4\times10^{12}\,$at/s, which corresponds to
the total initial flux of the thermal atoms of
$8.7\times10^{12}\,$at/s. It is assured here that
there are no losses of the atoms during their deceleration and
extraction from the slower.
 \begin{figure}
 \includegraphics[scale=0.8]{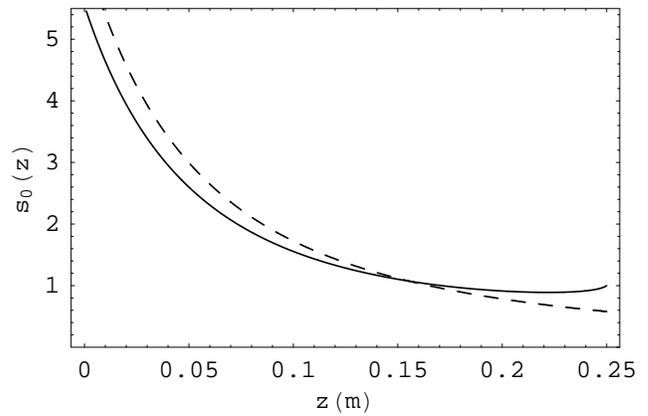}
 \caption{\label{}Spatial dependence of saturation parameter inside Zeeman slower 
in presence of 50\% absorption (solid line) and
without any absorption (dashed line).}
 \end{figure}

Figure\,1 shows
the calculated spatial distribution of the saturation parameter of
the atomic transition along the $z$-axis of the slower.
The solid curve shows that the large absorption of the slowing
light at the end of Zeeman slower, where the density of the slowed
atoms reaches its maximum, leads to slight decrease of the
saturation parameter, which can not be compensated by the
convergence of the laser beam. The dashed curve corresponds to the
case of absence of any absorption of the cooling light. Here the
smaller initial saturation parameter  $s_0(z_f)=0.58$ is used to
keep the capture velocity of the slower $v_0=420\,$m/s the same.
In the absence of absorption the saturation parameter is growing
monotonically due to convergence of the laser beam. Comparison of the
derivatives of these curves shows that the absorption
takes place mostly in the end half of the Zeeman slower, which is
explained by higher density of atoms and lower saturation
parameter at this part of the slower. 
 \begin{figure}
 \includegraphics[scale=0.7]{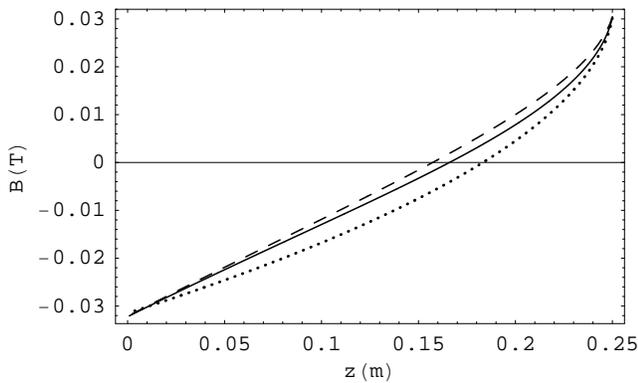}
 \caption{\label{}Spatial distribution of magnetic field inside Zeeman slower in presence of 50\% absorption and convergent laser beam (solid line); for convergent laser beam and no absorption (dashed line); 
collimated laser beam and no absorption (dotted line).}
 \end{figure}
Figure 2 shows the optimal spatial
distribution of the magnetic field along the Zeeman slower,
calculated from the formulas (1-8). The solid curve corresponds to
the 50\% absorption of the laser beam and the other parameters as
above. The dashed curve shows the case, when the absorption is
absent and the initial saturation parameter $s_0(z_f)=0.58$.
The dotted curve corresponds to a uniform deceleration
of atoms in a laser beam of constant diameter and constant
saturation parameter $s_0(z)=1.5$ in absence of absorption.
\begin{figure}
 \includegraphics[scale=0.8]{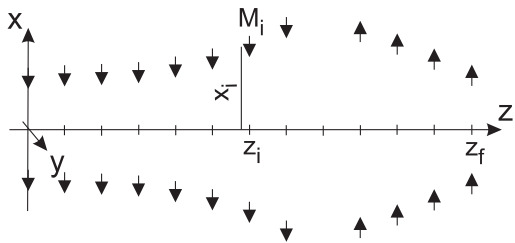}
 \caption{\label{}Schematic structure of Zeeman slower based on magnetic dipoles.}
 \end{figure}

\section{Zeeman slower based on magnetic dipoles}
It is proposed to use a Zeeman slower based on a quasiperiodic
array of small permanent magnets as is shown in fig.\,3. These
magnets can be modeled as point-like magnetic dipoles (MD).
There are two different possible configurations of such a slower,
the transverse, when the dipoles are oriented perpendicularly to the
axis of the slower (as in fig.\,3) or the longitudinal one, when
the dipoles are oriented along the axis and the resulted magnetic
field of the slower is also directed along that axis.

To understand the basic difference between the two configurations,
the field of a single magnetic dipole has to be considered first. For a
magnetic dipole placed at the origin of a Cartesian system of
coordinates and oriented along the x-axis, its magnetic field is
described by the formulas
\begin{eqnarray}
B_x=\frac{\mu_0 M}{4 \pi} \left[ \frac{2x^2-y^2-z^2}{r^5} \right] \nonumber \\
B_y=\frac{\mu_0 M}{4 \pi} \left[ \frac{3xy}{r^5} \right] \nonumber \\
B_z=\frac{\mu_0 M}{4 \pi} \left[ \frac{3xz}{r^5} \right],
\end{eqnarray}
where $r=\sqrt{x^2+y^2+z^2}$ and $M$ is the magnetic moment of the dipole.
Let us consider now a distribution of the magnetic field along
the $z$-axis for a single MD placed at $x=R, y=0, z=0$, while it is oriented
perpendicular or parallel to the $z$-axis.
The corresponding distributions of the proper components of magnetic field are given by
\begin{eqnarray}
B_{x}(z)=\frac{\mu_0 M}{4 \pi} \left[ \frac{2R^2-z^2}{(R^2+z^2)^{5/2}} \right] \nonumber \\
B_{z}(z)=\frac{\mu_0 M}{4 \pi} \left[ \frac{2z^2-R^2}{(R^2+z^2)^{5/2}} \right].
\end{eqnarray}
For a symmetric distribution of the magnetic dipoles around the
$z$-axis, all other components of the magnetic field at this axis are equal to zero.
Figure 4 shows the corresponding distributions of $B_{x}(z)$ (solid line) and
$B_{z}(z)$ (dashed line) for a MD with magnetic moment
$M=1.86\,$Am$^2$ and $R=3\,$cm.
\begin{figure}
 \includegraphics[scale=0.8]{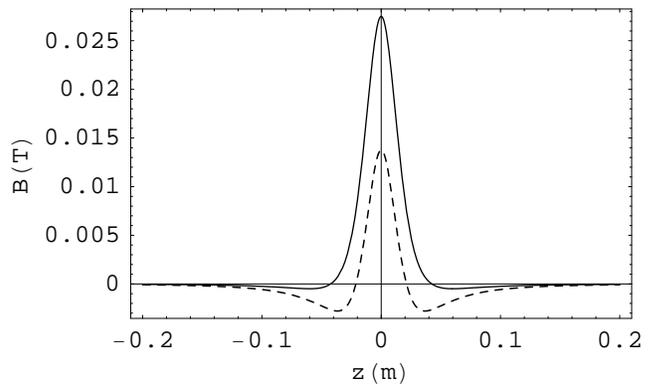}
 \caption{\label{}Spatial distribution of magnetic field of a point-like magnetic dipole. The distribution along the $z$-axis for a magnetic dipole oriented perpendicularly to the axis (solid line).  
The distribution for a dipole moment oriented parallel to the $z$-axis (dashed line).}
 \end{figure}
 
For the perpendicular orientation of the MD, its field along the $z$-axis
is mostly of the same sign, while for the
parallel orientation (dashed line), the maximum amplitude of the
field is twice smaller and its amplitude on the wings of the
distribution, where the field has an opposite sign, is comparable to
its maximum value. From the eq.\,(11) it follows that for
perpendicular MD its field turns to zero at $z=\sqrt{2}R$, while
for parallel orientation of the MD it happens at $z=R/\sqrt{2}$.
Therefore, the longitudinal configuration of the slower is less
preferable, because it demands much stronger magnets with smaller
spacing between them. An additional problem of the longitudinal
MD-Zeeman slower is that magnetic screening of it leads to further
decrease of its magnetic field.

For an array of magnets the uniformity of the resultant magnetic
field depends on the distances between the magnets. To
study this, an infinite periodic array of equidistant
($x=\pm R$) pairs of transverse MD separated along $z$-axis by a
constant interval $dz$ has been calculated. It was found that the
relative amplitude of the ripples of the magnetic field on
$z$-axis is about 1\% for a spacing
$dz=0.65 R$ and it decreases rapidly with further decrease of
$dz$.

The schematics of the transverse MD-Zeeman slower for Sr
atoms is shown in fig.\,3. It is designed to produce a magnetic
field, which is changing its sign, as it is shown in fig.\,2. The
slower consists of $n=13$ sections separated from each other by
the same interval $dz=2.083\,$cm. Each $i$-th section of the
slower consists of two MD of the same direction and placed
symmetrically with respect to the $z$-axis at $x=\pm x_i$. In the
first 8 sections the direction of the magnetic dipoles is opposite
to the $x$-axis and for the last four sections the dipoles are
directed along the axis. The 9th section has no
magnetic dipoles in it. To produce the desired spatial
distribution of the magnetic field the transverse distances $x_i$
of the magnetic dipoles have to be chosen properly. To find the
right distances of the magnetic dipoles a system of nonlinear
equations, which relates the magnitude of the magnetic field at
$n$ selected points on the $z$-axis of the slower to the sum of
the partial magnetic fields produced by $n$ sections of the
slower, can be solved. On the other hand, it was found that the
local field of the slower at $z=z_i$ is mostly determined by the
closest magnets of the $i$-th section of the slower and it is easy
to fit the target value of the local field by gradual change of
the corresponding pair of magnets.
\begin{figure}
 \includegraphics[scale=0.8]{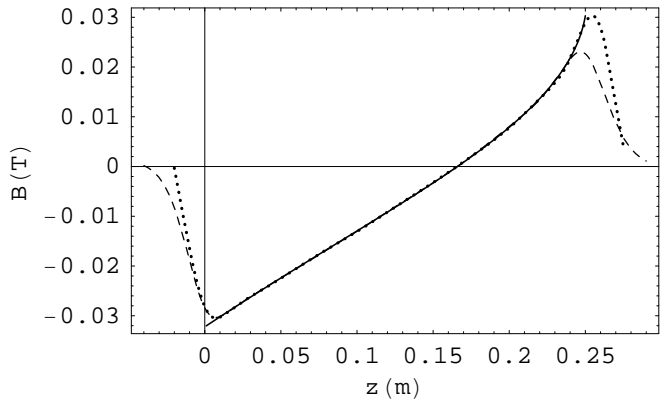}
 \caption{\label{}Spatial distribution of magnetic field inside the Zeeman
slower. The computed optimal field distribution (solid line). The
field produced by an array of transverse MDs (dashed line). The
field produced by the array of magnetic dipoles surrounded with a
magnetic shield (dotted line).}
 \end{figure}

An example of such a fit is shown in fig.\,5. The solid curve
shows the optimal magnetic field, which is the same as in fig.\,2.
The dashed curve shows the computed field produced by a transverse
MD-Zeeman slower, which consists of 13 equidistant section, 12 of which
include pairs of MD with magnetic moment $M=1.86\,$Am$^2$. The
dotted curve shows the distribution of the magnetic field of the
slower surrounded by two ideal magnetic shields, placed at $z=-2\,$cm
and $z=27.7\,$cm. To model the influence of the magnetic shields,
the images of the real magnets were added at proper distances from
both sides of the slower. As far as the images of the transverse magnetic dipoles
have an opposite sign, the magnetic shields should not be placed
too close to the ends of the slower. It was found also that the best
fit of the target field distribution at the end of the slower,
where it has maximal slope steepness, is achieved when the step between the
last two sections of the slower is increased to $2.7\,$cm. Therefore,
the axial position of the last section of the slower was taken to be
$z_{13}=25.7\,$cm. The corresponding most optimal distances of the
MD from the $z$-axis are shown in fig.\,6.
 \begin{figure}
 \includegraphics[scale=0.8]{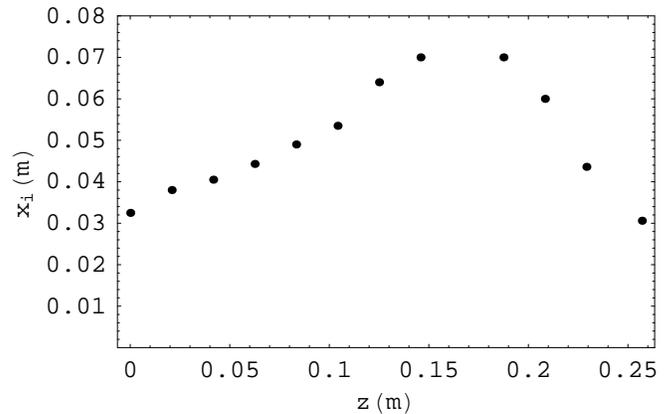}
 \caption{\label{}Distances of the individual magnetic dipoles from the axis of Zeeman slower.}
 \end{figure}

To calculate the magnetic field distribution of the MD-Zeeman
slower in the $x0y$-plane one needs to take into account all
three components of the magnetic field $B_x$, $B_y$ and $B_z$,
which define the total amplitude of the magnetic field
$B(x,y,z)=\sqrt{B_x^2+B_y^2+B_z^2}$ responsible for the local
Zeeman shift of atomic magnetic states. The transverse variation
of the field is the strongest at the input and output edges of
Zeeman slower, where the distances of the magnets from the axis of
the slower are the smallest. The corresponding distributions of
the magnetic field amplitude near the axis of symmetry of the
slower along the $x$-axis (dashed line) and the $y$-axis (solid
line), taken at the output plane of the slower at $z=25.7\,$cm,
are shown in fig.\,7. In the central region
of the slower, where the distances of the magnets from the
$z$-axes of the slower are larger, the transverse variation of
the field is much smaller.
\begin{figure}
 \includegraphics[scale=0.75]{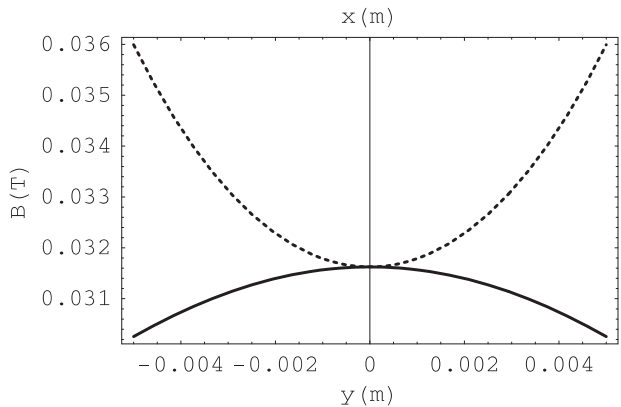}
 \caption{\label{}Transverse spatial distribution of magnetic field along the $x$-axis 
(dashed line) and the $y$-axis (solid line)
at the output plave of the MD-Zeeman slower at $z=25.7\,$cm.}
 \end{figure}

\begin{figure}
 \includegraphics[scale=0.66]{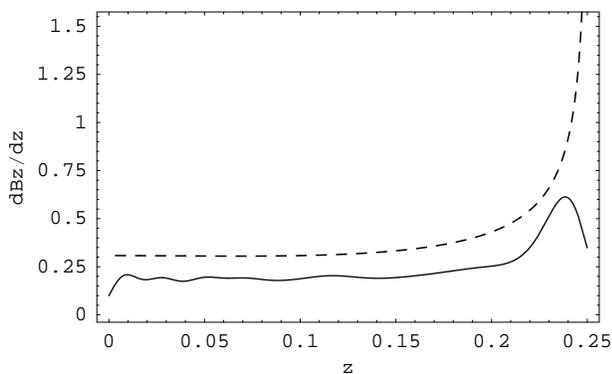}
 \caption{\label{}Spatial distribution of maximum gradient of magnetic field inside the slower, which provides
continuous deceleration of atoms (dashed line). The computed distribution of the magnetic field
gradient of the magnetic field of the considered MD-Zeeman slower (solid line).}
 \end{figure}

The well known limit on the maximal slope
steepness of the magnetic field of a Zeeman slower is given by the ratio
\begin{equation}
\frac{dB(z)}{d z}<\frac{\hbar k a_{max}(z)}{\mu_B v(z)}.
\end{equation}
This formula is derived from the condition
that the local deceleration rate $a(z)$ can't exceed $a_{max}(z)$.
This condition is true, but not complete.
If the local deceleration rate $a(z)$ exceeds the maximum value
$a_{max}(z)$, but only for a short period of time $\delta t$, the
atoms can still be further decelerated along the stable
velocity-trajectory of a Zeeman slower. The maximal duration of
this time can be estimated from the ratio $(a(z)-a_{max}(z))
\delta t=v_{res}(z)-v(z)$, which means that the exceeding
acceleration should be able to accelerate an atom during the time
$\delta t$ from its equilibrium velocity to the resonant velocity,
at which atom is lost from the further deceleration process. The
dashed line in fig.\,8 shows the spatial distribution of the
accepted maximal local gradient of the magnetic field, computed
numerically from the eq.\,12 for a decelerated atom.
The solid line represents the actual gradient of the
magnetic field, produced by an array of magnetic dipoles of the
MD-Zeeman slower.
 \begin{figure}
 \includegraphics[scale=0.7]{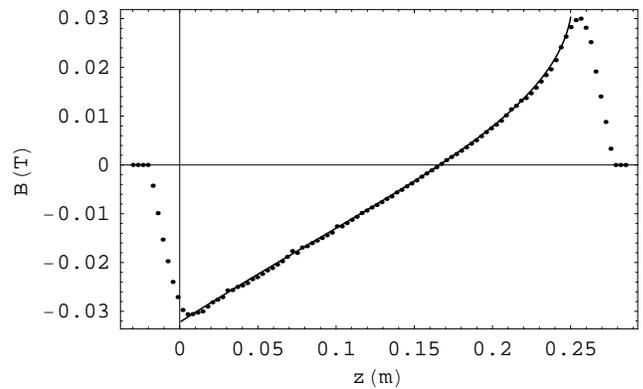}
 \caption{\label{}The optimal magnetic field distribution (solid line) and the
simulated field distribution of the slower composed of finite-size
neodymium magnets surrounded with an iron magnetic shield.}
 \end{figure}

The magnetic moment of a real magnet is determined as $M=B_i
V/\mu_0$, where $B_i$ is intrinsic induction and $V$ is the total
volume of the magnet. The value $M=1.86\,$Am$^2$ in the examples
considered above corresponds to a cylindrical neodymium magnet
with $B_i=1.1\,$T, diameter $1.5\,$cm and height $1.2\,$cm. Figure
9 shows a precise calculation of the MD-Zeeman slower based on
such finite-size magnets performed with the finite element
analysis software COMSOL. In these calculations the distances of
the centers of the magnets from the axis of the slower were taken
exactly the same as in the MD model described above. The positions
of the iron magnetic shields were also the same as in fig.\,6.

\section{Outlook}

The mechanical construction of such a slower can be rather simple
and inexpensive. It can be a box made of iron or some other
material of high magnetic permeability, which serves as a frame
for a two sets of screws with cylindrical magnets attached to
their bottoms. The box is used both for holding the screws and for
magnetic screening of the slower. In such a construction the distances of the
individual magnets from the axis of the slower can be adjusted by
rotation of the corresponding screws, without changing their axial
positions.

In the considered MD-Zeeman slower all the magnets of the same size were used.
In practice it make sense to use in the centre of the slower smaller magnets
placed closer to the axis of the slower, which will make the whole construction
more compact.

The maximum flux of cold atoms produced by a Zeeman slower is
usually decreased by a transverse heating of the slowed atoms
via the spontaneous scattering of the cooling light. This heating can be
partially compensated by introducing
additional sections, where the transverse motion of atoms is
cooled down with additional transverse laser beams. This makes
possible to increase the total flux of the cold atoms typically by
one order of the magnitude or more. Such a transverse cooling of
the atoms can be performed before \cite{Slowe}, after \cite{Lison}
or in between the two sections \cite{Joffe} of a standard Zeeman
coil magnet. The problem is that the access to the atoms in
the transverse direction is normally completely blocked by the coils.
In a transverse MD-Zeeman slower the two arrays of compact magnets are placed
from the two sides of the atomic beam and it is easy to get access
to it on the whole length of the atomic beam. Therefore, for a
MD-Zeeman slower it is possible to cool atoms transversely along the
whole length of the slower.

Finally, a few words on the operation of the transverse MD-Zeeman
slower. A transverse slower uses linearly
polarized light, which can be presented as a linear superposition of
two ($\sigma^+$ and $\sigma^-$) circularly polarized components.
Therefore, only one-half of the total light intensity is
in resonance with the right Zeeman transition of the
decelerated atoms. An additional complication arises when a
Zeeman splitting of magnetic sublevels of the ground state of an
atom is comparable to the natural linewidth of the atomic
transition. As it was recently shown in \cite{Balykin}, a
transverse Zeeman slower for Rb atoms demands quite some
additional light power to make it work in the presence of the optical
pumping between the split magnetic sublevels of the
ground state. Fortunately, the Zeeman splitting of the ground state
of the atoms from the earth-metal group is very small
\cite{Lemonde2}. That is why such a transverse MD-Zeeman slower is very promising for
use in an optical atomic clock based on such atoms.

\section{Conclusion}

A standard theory of a Zeeman slower is extended to a case of a nonuniform intensity
distribution of the cooling light.
A way to include the absorption of the cooling light into the calculation of the
optimal magnetic field of a Zeeman slower is described.
It is shown that in that case the numerical simulation of the acceleration of atoms starting from their final 
position in the slower and calculating in the backwards direction is preferable. A new design parameter $\epsilon$ instead
of the standard one $\eta$ parameter is proposed. The main advantage of such a definition is that the
most optimal cooling of the decelerated atoms is reached at exactly $\epsilon=0.75$.

A transverse Zeeman slower composed of an array of discrete compact magnets is proposed and a
simple model of such a slower is developed.
As an example, a compact transverse Zeeman slower for Sr atoms has been calculated in
presence of 50\% absorption of the cooling light. The validity of the simple MD model of the slower
for a slower composed of finite-size magnets has been confirmed with precise numerical calculations.

\acknowledgments
Many thanks to Anne Curtis and Christopher Foot for the valuable comments.
This work was funded by the UK National Measurement System Directorate of the Department of Trade and Industry.

\end{document}